\documentclass[a4paper,12pt]{article}

\usepackage{amssymb,latexsym,amsmath}
\usepackage{bbold}

\usepackage{cite}

\usepackage{fullpage}

\usepackage[pdftex]{graphicx}

\makeatletter \@addtoreset{equation}{section} 
\makeatother

\begin{document} 
\begin{titlepage}
	\thispagestyle{empty} 
	\begin{flushright}
		\hfill{DFPD-2014/TH/17}
	\end{flushright}
	
	\vspace{35pt} 
	\begin{center}
		{\LARGE\textbf{Uplifting non-compact \\[3mm] gauged supergravities}}
		
		\vspace{60pt}
		
		{Walter H. Baron and Gianguido Dall'Agata}
		
		\vspace{30pt}
		
		{\it Dipartimento di Fisica e Astronomia ``Galileo Galilei'' \\
		Universit\`a di Padova, Via Marzolo 8, 35131 Padova, Italy}
		
		\vspace{20pt}
		
		\emph{and}
		
		\vspace{20pt}
		
		{\it INFN, Sezione di Padova \\
		Via Marzolo 8, 35131 Padova, Italy}
		
		\vspace{40pt}
		
		{ABSTRACT} 
	\end{center}
	
	\vspace{10pt}
	
We provide the M-theory uplift of de Sitter vacua of SO(5,3) and SO(4,4) gaugings of maximal supergravity in 4 dimensions. We find new non-compact backgrounds that are squashed hyperboloids with non-trivial flux for the 3-form potential. 
The uplift requires a new non-linear ansatz for the 11-dimensional metric and for the 3-form potential that reduces to the known one leading to the 7-sphere solution in the case of the SO(8) gauging.
\end{titlepage}

\baselineskip 6 mm

\section{Introduction} 
\label{sec:introduction}

Supergravity theories are generally believed to describe some low-energy approximation of string theory models.
There is, however, a huge landscape of 4-dimensional supergravity models whose higher-dimensional origin is not clear yet.
This is already true for the most constrained scenario of maximal supergravities, which we therefore decided to explore, looking for new general uplifting/reduction procedures.

The original $N=8$ ungauged supergravity model \cite{Cremmer:1978ds} can be directly obtained by reducing 11-dimensional supergravity on a torus \cite{Cremmer:1979up}.
However, there are infinite possible deformations compatible with maximal supersymmetry, which are obtained in the gauging process.
Using the embedding tensor formalism \cite{Nicolai:2000sc,deWit:2002vt} these deformations are encoded in a set of 912 parameters subject to a quadratic constraint.
The first and most notable example is the SO(8) gauging of de Wit and Nicolai \cite{deWit:1983gs}, which is the result of an M-theory reduction on the 7-dimensional sphere \cite{deWit:1986iy,Nicolai:2011cy}.
While it is somewhat easy to check that the linearized theory gives the correct equations and that the $N=8$ vacuum of the SO(8) theory has a spectrum that is contained in the general Kaluza--Klein analysis, it is much more difficult to construct the full uplift of the 4-dimensional degrees of freedom into the 11-dimensional ones  \cite{Nicolai:2011cy}.
Actually, only recently the uplift ansatz for the 3-form has been completed in \cite{deWit:2013ija} and this allowed for many non-trivial checks, including the uplift of various vacua with lower residual symmetry, obtained squashing the original $S^7$ geometry \cite{Godazgar:2013nma,Godazgar:2013pfa,Godazgar:2014eza}.

Much less is known if one looks at different and possibly non-compact gaugings.
Scherk--Schwarz reductions are the first example of a general procedure to obtain a large class of models with Minkowski vacua in 4 dimensions \cite{Cremmer:1979uq}.
The full non-linear uplift required some work and has the nice interpretation of a dimensional reduction on twisted tori \cite{Kaloper:1999yr}. 
Once more, only very recently the relation between the 4-dimensional models expressed in terms of the embedding tensor formalism and the 11-dimensional uplift has been obtained \cite{Godazgar:2013oba}.
From the 4-dimensional point of view some of these models correspond to instances in the general class of contractions of the SO($p,q$) gaugings, where $p+q =8$ \cite{Catino:2013ppa}.
For the maximally symmetric de Sitter vacua of the SO(5,3) and SO(4,4) models, Hull and Warner provided the 11-dimensional uplift \cite{Hull:1988jw} by means of a non-compact manifold, whose shape is that of an hyperboloid, but we still lack a general procedure that allows for the uplift of other vacua.
In fact an old approach for finding vacua of potentials whose scalars parametrize coset manifolds \cite{Li:1986tk} revealed to be extremely powerful in combination with the embedding tensor formalism \cite{DallAgata:2011aa}, allowing for the analytic discovery of many new solutions of compact and non-compact $N=8$ gaugings \cite{Dall'Agata:2012bb,Dall'Agata:2012sx,Borghese:2012qm,Borghese:2012zs,Kodama:2012hu,Borghese:2013dja,Catino:2013ppa,Dall'Agata:2014ita,Gallerati:2014xra}.
In this process, it was also discovered that the possible deformations for a given gauge group, including maximal ones like SO(8) and SO($p,q$), are infinite \cite{Dall'Agata:2012bb} and their vacuum structure changes according to the value of the deformation parameter.
This hugely increased both the number of theories and the number of vacua of maximal supergravity that are not yet understood as the result of some string theory reduction and make even more compelling a better understanding of general reduction processes.

In this paper we make a small but significant step in this direction, providing an ansatz for the 11-dimensional metric and 3-form potential that allows to uplift all the known vacua of the regular SO(5,3) and SO(4,4) gaugings.
This includes the vacua of \cite{Hull:1988jw}, but also the new one discovered in \cite{Dall'Agata:2012sx}.
We are going to show that, in order to solve the 11-dimensional equations of motion, one has to slightly, but crucially, modify the ansatz of de Wit and Nicolai \cite{deWit:2013ija}, taking into account the transformation properties of the Killing vectors generating the symmetry group of the gauged theory under SL($8,{\mathbb R}$).
In detail, we propose the following ansatz for the metric
\begin{equation}
	\Delta^{-1}(x,y) g^{mn}(x,y) = K^m{}_{AB}(y) \, K^n{}_{CD}(y) \,{\cal V}^{AB\,ij}(x)\, {\cal V}^{CD}{}_{ij}(x),\label{MAmixframe}
\end{equation}
and for the 3-form potential 
\begin{equation}
	A_{mnp}(x,y) = \frac{1}{\sqrt2} \, \Delta(x,y) \, g_{pq}(x,y) \, K_{mn}{}^{AB}(y)\,  K^q{}_{CD}(y)\,{\cal V}_{AB\,ij}(x)\,{\cal V}^{CD\,ij}(x).\label{FAmixframe}
\end{equation}
Here ${\cal V}$ are the coset representatives of the 4-dimensional scalar manifold, $K^m{}_{AB}$ are the Killing vectors of a maximally symmetric hyperbolic space and $K_{mn}{}^{AB}$ are their covariant derivatives.
The $A,B,\ldots$ indices are raised and lowered with the SO($p,q$) metric, namely $\eta = {\rm diag}\{\underbrace{1,\ldots,1}_p, \underbrace{-1,\ldots,-1}_q\}$.
These uplift formulas work for Freund--Rubin like reductions, where the 4-dimensional part of the solution is given by 
\begin{eqnarray}
g_{\mu\nu}(x,y)=\Delta^{-1}(x,y)\,\tilde g_{\mu\nu}(x)\,,
\label{4Dmetric}
\end{eqnarray}
\begin{eqnarray}
F_{\mu\nu\rho\sigma}=f_{FR}\,\epsilon_{\mu\nu\rho\sigma}\,,
\label{4DF}
\end{eqnarray}
where $\epsilon_{\mu\nu\rho\sigma}$ is the four dimensional volume form associated to the de Sitter (Anti de Sitter in the case $q=0$) background metric $\tilde g_{\mu\nu}$, and there are no contributions with mixed internal and external indices, neither for the metric nor for the 4 form.
For $q=0$, we reproduce the ansatz of \cite{deWit:2013ija} for the SO(8) gauging, but crucial signs differences arise in the more general case of the SO($p,q$) gaugings.

In the following, after a brief review of the necessary ingredients to build the new solution, we first derive our new ansatz for the metric and 3-form, we then test it against the solutions of \cite{Hull:1988jw} and finally use it to construct a new solution of 11-dimensional supergravity, corresponding to the uplift of the SO(3) $\times$ SO(3) invariant de Sitter vacuum of the SO(4,4) theory \cite{Dall'Agata:2012sx}.
The resulting metric is a squashed hyperboloid, which also supports a non-trivial flux for the 4-form field strength.

The last part of our paper is devoted to the relation of our results with (some notion of) generalized geometry.
The ideal objective of this kind of research is the development of a general procedure to find stringy reductions to 4 dimensions producing effective theories or consistent truncations corresponding to arbitrary gaugings of (maximal) supergravities, including all its vacua as special cases.
Among several attempts in this direction, inspired by \cite{Coimbra:2011ky,Coimbra:2012af,Aldazabal:2013,Aldazabal:2013b,Hohm:2013,Hohm:2013b,Hohm:2013c,Lee:2014mla,Baron:2014yua}, we will show that our ansatz follows also quite naturally if one assumes that the 11-dimensional degrees of freedom are encoded in a generalized 56-dimensional vielbein
\begin{equation}
	\label{genviel}
	{\cal E}_{\mathbb A}{}^{\mathbb M}(x,y) = U_{\mathbb A}{}^{\mathbb B}(x) E_{\mathbb B}{}^{\mathbb M}(y), 
\end{equation}
where $E$ produces the algebra structure when acted upon by a generalized Lie derivative (to be defined later)
\begin{equation}
	L_{E_{\mathbb A}} E_{\mathbb B} = X_{{\mathbb A}{\mathbb B}}{}^{\mathbb C} E_{\mathbb C}\label{LieDerE}
\end{equation}
and $U(x)$ encodes the coset representative of the E$_{7(7)}$/SU(8) scalar manifold.
This obviously cannot be the final ansatz, and, as we will discuss later, does not seem to be efficient enough to include all gaugings, but it includes all cases discussed so far and may be used to further extend our analysis.


\section{The new ansatz for the uplift} 
\label{sec:the new ansatz for the uplift}

\subsection{Preliminaries} 
\label{sub:preliminaries}

In this work we are concerned with the SO($p,q$) gaugings obtained as electric subgroups of the SL($8,{\mathbb R}$) group of maximal supergravity.
In the SL($8,{\mathbb R}$) basis, the 133 generators of the U-duality group E$_{7(7)}$ can be written as
\begin{equation}\label{e7 gen in sl8 basis}
  [t_\alpha]_{\mathbb A}{}^{\mathbb B} \equiv \left(\begin{array}{cc}
      \Lambda_{AB}{}^{CD} & \Sigma_{ABCD} \\
      \star\Sigma^{ABCD} & \Lambda'^{AB}{}_{CD} 
  \end{array}
  \right),
\end{equation}
according to the following decompositions of the fundamental and adjoint representations of ${\mathfrak e}_{7(7)}$:
\begin{equation}\label{sl8 decomposition}
 {\bf 56} \longrightarrow {\bf 28} + {\bf 28'},\quad
 {\bf 133} \longrightarrow {\bf 63} + {\bf 70}.
\end{equation}
Here $\Lambda_{AB}{}^{CD} \equiv 2 \Lambda_{[A}{}^{[C} \delta_{B]}{}^{D]}$ and
$\Lambda' = -\Lambda^T$, with  $A,B,\ldots$ denoting the $\bf8$ and $\bf8'$ representations of SL(8,${\mathbb R}$),
while $\Sigma_{ABCD}$ and $\star\Sigma^{ABCD}$ denote (anti) selfdual real forms.
Coset representatives are constructed by exponentiation
\begin{equation}\label{sl8coset}
	{\cal U}(\phi) = \left(\begin{array}{cc}
	{\cal U}_{AB}{}^{\underline{cd}} & {\cal U}_{AB\,\underline{cd}} \\[2mm]
	{\cal U}^{AB\,\underline{cd}} & {\cal U}^{AB}{}_{\underline{cd}} 
	\end{array}\right) = 
  \left(\begin{array}{cc}
	u^{\underline{cd}}{}_{AB} & -v_{\underline{cd}\,AB} \\[2mm]
	-v^{\underline{cd}\,AB} & u_{\underline{cd}}{}^{AB} 
	\end{array}\right)=\exp\left(\phi^{\hat{\alpha}}t_{\hat{\alpha}}\right), \quad t_{\hat \alpha} \in \mathfrak{e}_{7(7)}\setminus\mathfrak{su}(8),
\end{equation}
where $\phi^{\hat \alpha}$ parametrize the scalar manifold. 
Lower case indices are acted upon by local SU(8) transformations, while capital indices are acted upon by rigid E$_{7(7)}$ transformations.
Given the structure of the scalar manifold and the fact that the SU(8) subgroup of E$_{7(7)}$ is the R-symmetry group of the theory, it is useful to rewrite (\ref{sl8coset}) in the complex SU(8) basis.
The change of basis is performed by using real chiral $\Gamma$ matrices that interpolate between the two sets of indices, using the triality property of the common SO(8) subgroup.
Introducing
\begin{equation}\label{cayley matrix and complex basis}
 S \equiv 
  \frac{1}{4\sqrt2}\; (\Gamma^{ij})_{AB} \otimes
 \left(\begin{array}{cc}
     1 &  i \\
  	 1 & -i \\
	 \end{array}\right)
,
\end{equation}
we can transform the 56-dimensional vector in the SL($8,{\mathbb R}$) basis to the corresponding one in the SU(8) basis:
\begin{equation}
	\left(\begin{array}{c}
	z_{ij} \\
	z^{ij} = (z_{ij})^*
	\end{array}\right) = S \left(\begin{array}{c}
	x_{AB} \\ y^{AB}
	\end{array}\right).
\end{equation}
Since we use the double index notation, normalizations are fixed so that $(SS^{-1})^{ij}{}_{kl} = \delta^{ij}_{kl}$.
We therefore introduce two further representations of the coset representative.
The first one is the SU(8) representation obtained by $U_{SU_8} = S {\cal U} S^{-1}$.
As noted in \cite{deWit:2007mt}, the resulting matrix is the inverse\footnote{We recall that in this representation $U_{SU_8}^\dagger \omega U_{SU_8} = \omega$, with $\omega$ defined in (\ref{omegadef}).} of the standard vielbein representation of \cite{deWit:1983gs}, which is expressed in terms of the SU(8) tensors $(u_{ij}{}^{KL})^* = u^{ij}{}_{KL}$ and $(v_{ijKL})^* = v^{ijKL}$ as
\begin{equation}\label{coset representatives su basis}
	 U_{SU_8} = \left(\begin{array}{cc}
	 U_{IJ}{}^{kl} & U_{IJ\, kl} \\
	 U^{IJ\, kl} & U^{IJ}{}_{kl}
 	 \end{array}\right) = 
	 \left(\begin{array}{cc}
	  u^{kl}{}_{IJ} & -v_{klIJ} \\
	 -v^{klIJ} & u_{kl}{}^{IJ}
 	 \end{array}\right).
\end{equation}
The indices $i,j,\ldots$ and $I,J,\ldots$ take values from 1 to 8 and the distinction is related to the different actions of the coset generators on the representatives.
Again lower case indices are local under SU(8) transformations, while capital indices are rigid under E$_{7(7)}$.
It is also very useful to introduce the mixed basis representative
\begin{equation}\label{coset representatives mixed basis}
{\cal V} = 
	 \left(\begin{array}{cc}
	 {\cal V}_{AB}{}^{kl} & {\cal V}_{AB\, kl} \\
	 {\cal V}^{AB\, kl} & {\cal V}^{AB}{}_{kl}
 	 \end{array}\right) = {\cal U} S^{-1},
\end{equation}
which implies
\begin{eqnarray}
	{\cal V}_{AB}{}^{kl} &=& ({\cal V}_{AB\, kl})^* = \frac{1}{4 \sqrt2} (\Gamma^{IJ})_{AB} \left(u^{kl}{}_{IJ} - v^{klIJ}\right), \\
	{\cal V}^{ABkl} &=& ({\cal V}^{AB}{}_{kl})^* = -i \frac{1}{4 \sqrt2} (\Gamma^{IJ})_{AB} \left(u^{kl}{}_{IJ} + v^{klIJ}\right).
\end{eqnarray}
This last representation is especially useful if we want to extract a correct ansatz for the uplift of the non-compact 4-dimensional models, following the same line of reasoning as the one presented in \cite{deWit:2013ija}.
We recall that this last representation satisfies \cite{deWit:2007mt}
\begin{eqnarray}
	{\cal V}_{\mathbb A}{}^{ij} {\cal V}_{{\mathbb B}\,ij} - {\cal V}_{{\mathbb A}\,ij}{\cal V}_{\mathbb B}{}^{ij} &=& i \, \Omega_{\mathbb{AB}}, \\
	\Omega^{\mathbb{AB}}\, {\cal V}_{\mathbb A}{}^{ij}{\cal V}_{{\mathbb B}\,kl} &=& i \, \delta^{ij}_{kl}, \\
	\Omega^{\mathbb{AB}} \, {\cal V}_{\mathbb A}{}^{ij}{\cal V}_{\mathbb B}{}^{kl} &=& 0,
\end{eqnarray}
or equivalently  
\begin{eqnarray}
{\cal V} \omega{\cal V}^\dagger	&=& \, \Omega\;,
\end{eqnarray}
with
\begin{eqnarray}
\Omega=\left(\begin{matrix}0&{\mathbb 1}_{28}\cr-{\mathbb 1}_{28}&0\end{matrix}\right)\;,\qquad
\omega=i\, \left(\begin{matrix}-{\mathbb 1}_{28}&0\cr0&{\mathbb 1}_{28}\end{matrix}\right), \label{omegadef}
\end{eqnarray}
which means $\Omega_{AB}{}^{CD} = - \Omega^{CD}{}_{AB} = \delta_{AB}^{CD}$.


\subsection{Towards the new ansatz} 
\label{sub:towards_the_new_ansatz}

In order to extract a meaningful ansatz for the 11-dimensional metric and 3-form potential, we compare the 4-dimensional supersymmetry transformation rules with the 11-dimensional ones for the relevant field components.
Following \cite{deWit:2013ija,Godazgar:2013dma}, this is going to lead to the definition of a set of generalized vielbeins, whose properties include a natural ansatz for the metric and 3-form.

The starting point of the analysis is given by the supersymmetry transformation rules of the 4-dimensional gauge vectors.
In the SL($8,{\mathbb R}$) basis these are \cite{deWit:2007mt}
\begin{equation}
	\delta A_\mu^{{\mathbb A}} = -i\, \Omega^{\mathbb{AB}} {\cal V}_{\mathbb B}{}^{ij} \left(\overline \epsilon^k \gamma_\mu \chi_{ijk}+ 2 \sqrt2\, \overline \epsilon_i \psi_{\mu j}\right) + \rm{h.c.}\,,
\end{equation}
which can also be expressed in terms of the scalar fields as
\begin{eqnarray}
	\delta A_{\mu}^{AB} &=& \frac{1}{4 \sqrt2} (\Gamma^{IJ})_{AB} \left(u^{ij}{}_{IJ} + v^{ij\,IJ}\right)  \left(\overline \epsilon^k \gamma_\mu \chi_{ijk}+ 2 \sqrt2\, \overline \epsilon_i \psi_{\mu j}\right) + \rm{h.c.}\,, \label{susyvec}\\
	\delta A_{\mu\,AB} &=& -\frac{i}{4 \sqrt2}\,  (\Gamma^{IJ})_{AB} \left(u^{ij}{}_{IJ} - v^{ij\,IJ}\right) \left(\overline \epsilon^k \gamma_\mu \chi_{ijk}+ 2 \sqrt2\, \overline \epsilon_i \psi_{\mu j}\right) + \rm{h.c.}\,. \label{susydualvec}
\end{eqnarray}
This differs from the expressions in \cite{deWit:1983gs,deWit:2013ija} by an invertible real matrix constructed with the antisymmetric $\Gamma$ matrices $(\Gamma^{IJ})_{AB}$, which, however can always be reabsorbed in the vectors by a simple field redefinition.
The electric gaugings we consider have $A_\mu^{AB}$ as the physical vector fields, while the $A_{\mu\,AB}$ are the dual potentials, which are going to disappear from the final 4-dimensional lagrangian.

In accordance with the standard non-linear ansatz for the SO(8) gauging \cite{deWit:2013ija}, we expect that $A_\mu^{AB}$ will be contained in the reduction ansatz of the 11-dimensional metric, while $A_{\mu\, AB}$ will be contained in the reduction ansatz of the 11-dimensional 3-form potential.
In detail, the vector gauge fields appearing in the 11-dimensional metric components $B_\mu^m$ are multiplied by the 28 Killing vectors of the sphere (or their generalizations for more general gaugings), so that 
\begin{equation}\label{Bmum}
	B_\mu{}^m(x,y) = -\frac12\, K^m_{AB}(y)\, A_\mu{}^{AB}(x),
\end{equation}
while the dual vector fields, appearing in the 3-form components $A_{\mu mn}$ are multiplied by the covariant derivatives of the Killing vectors (we will give momentarily the precise definition of these quantities for general gaugings in SL(8,${\mathbb R}$))
\begin{equation}\label{Bmumn}
	B_{\mu mn}(x,y) = -\frac{1}{2\sqrt2}\,\, K_{mn}{}^{AB}(y)\, A_{\mu\,AB}(x),
\end{equation}
where $B_{\mu mn}(x,y) \equiv A_{\mu mn} - B_\mu^p A_{mnp}$.
A crucial point in this discussion is the fact that the 11-dimensional quantities should be invariant under global transformations of the SO($p,q$) group and therefore the transformation properties of the Killing vectors and their derivatives, appearing in (\ref{Bmum}) and (\ref{Bmumn}), should be opposite to those of the vector fields, so that the final quantity is invariant.
Since indices in the $\mathbf{28}$ and $\mathbf{28}'$ are related to each other by means of the invariant metric of SO($p,q$), which we call $\eta$, $K^m{}_{AB} = K^{m\,AB}$ and $K_{mn\,AB} = K_{mn}{}^{AB}$ only in the special case of the SO(8) gauging, while in general
\begin{equation}
	K^m{}_{AB} = \eta_{AC} K^{m\,CD} \eta_{DB}.
\end{equation}
The supersymmetry variation of (\ref{Bmum}) and (\ref{Bmumn}) should lead to the supersymmetry transformation of the vectors and dual vector fields in (\ref{susyvec}) and (\ref{susydualvec}), with a multiplicative factor in front of the 4-dimensional spinor fields that is proportional to the generalized vielbein:
\begin{equation}
	\delta B_\mu{}^m(x,y) = \frac{\sqrt{2}}{8}\, e^{m \, ij}(x,y) \left(\overline \epsilon^k \gamma_\mu \chi_{ijk}+ 2 \sqrt2\, \overline \epsilon_i \psi_{\mu j}\right)(x) + \rm{h.c.}\,,
\end{equation}
\begin{equation}
	\delta B_{\mu mn}(x,y) = \frac{\sqrt{2}}{8}\, e_{mn}{}^{ij}(x,y) \left(\overline \epsilon^k \gamma_\mu \chi_{ijk}+ 2 \sqrt2\, \overline \epsilon_i \psi_{\mu j}\right)(x) + \rm{h.c.}\,.
\end{equation}
In these variations we used 4-dimensional spinors, but one can recover the 11-dimensional ones, assuming that the reduction ansatz is factorized thanks to the Killing spinors.
This fixes for us the form of the generalized vielbeins by coupling the appropriate Killing vectors with lower indices to the mixed coset representatives with upper SL(8,${\mathbb R}$) indices and the derivatives of the Killing vectors with upper indices to the mixed coset representatives with lower SL(8,${\mathbb R}$) indices:
\begin{eqnarray}
	e^m{}_{ij}(x,y) &=& 2\sqrt2\,i \, K^m{}_{AB}(y) \,{\cal V}^{AB}{}_{ij}(x), \label{emu}\\[2mm]
	e^{m\,ij}(x,y) &=& -2\sqrt2\,i \, K^m{}_{AB}(y) \,{\cal V}^{AB\, ij}(x), \label{emd} \\[2mm]
	e_{mn\,ij}(x,y) &=& -2i\, K_{mn}{}^{AB}(y) \,{\cal V}_{AB\,ij}(x), 	\label{emnd}\\[2mm]
	e_{mn}{}^{ij}(x,y) &=& 2i \, K_{mn}{}^{AB}(y) \,{\cal V}_{AB}{}^{ij}(x)\,. \label{emnu} 
\end{eqnarray}

These expressions coincide with the ones provided in \cite{deWit:2013ija} if one assumes that the Killing vectors and vector fields in \cite{deWit:2013ija} are related to ours by proper contractions with the $\Gamma$ matrices and recall that the gauge group is SO(8).
However, we can now use this as a starting point to describe more general gaugings contained in SL($8,{\mathbb R}$).
For this reason, we now describe the relevant geometric quantities for an arbitrary SO($p,q$) electric gauging.


\subsection{SO($p,q$) gaugings and hyperbolic geometry} 
\label{sub:so_p_q_gaugings_and_hyperbolic_geometry}

The starting point is the lorentzian hyperbolic space ${\cal H}_L^{p,q}$ (we assume $p+q = 8$ in the following), which can be embedded in ${\mathbb R}^{p,q}$ by means of the quadratic constraint
\begin{equation}\label{HypEmbed}
	\eta_{AB} X^A X^B = R^2,
\end{equation}
where $\eta$ is a constant metric with $p$ diagonal entries equal to +1 and $q$ diagonal entries equal to -1.
Geometrically, we can think of this space as a hyperbolic fibration of $S^{p-1} \times S^{q-1}$, and a useful parametrization follows by imposing

\begin{equation}
	\begin{array}{rcl}
	X^1 &=& R \cosh \psi\, \sin \theta_1 \ldots \sin \theta_{p-1}, \\[2mm]
	\ldots \\
	X^{p} &=& R \cosh \psi \, \cos \theta_1, \\[2mm]
	X^{p+1} &=& R \sinh \psi\, \sin \phi_1 \ldots \sin \phi_{q-1}, \\[2mm]
	\ldots \\
	X^{p+q} &=& R \sinh \psi \, \cos \phi_1. \\[3mm]
	\end{array}\label{paramer}
\end{equation}
If we imagine this hypersurface embedded in ${\mathbb R}^{p,q}$ endowed with the standard flat metric, the hypersurface itself inherits a metric with $(p,q)$ signature:
\begin{equation}\label{Lormet}
	ds^2=\eta_{AB}\; dX^A dX^B = \; \stackrel{\circ}{g}_{L\,mn}(y)\,dy^m dy^n.
\end{equation}
This metric is maximally symmetric and admits 28 Killing vectors generating the $\mathfrak{so}(p,q)$ algebra:
\begin{equation}\label{KillVec}
	K^{AB}= R^{-1}\,\left(X^A \eta^{BC}-X^B \eta^{AC}\right)\partial_C, 
\end{equation}
where $\partial_C$ must be understood as the derivative $\frac{\partial}{\partial X^C}$ constrained by (\ref{HypEmbed}). 
On the other hand, if we take the same constrained surface (\ref{HypEmbed}) embedded in ${\mathbb R}^8$, we obtain the euclidean hyperbolic metric 
\begin{equation}\label{euc}
	ds^2=\delta_{AB}\, dX^A dX^B =\, \stackrel{\circ}{g}_{E\,mn}(y)\,dy^m dy^n,
\end{equation}
whose isometry is restricted to the SO($p$) $\times$ SO($q$) group\footnote{While maximally symmetric spaces are unique \cite{Weinberg}, this is not the case for those with reduced isometry groups. 
In particular the Euclidean spaces with metric (\ref{euc}) and embedding
$(\tilde{X}^1)^2+\dots + (\tilde X^{p})^2-c^2\left((\tilde X^{p+1})^2+\dots +(\tilde X^{d})^2\right)=R^2$ also possess an isometry group SO$(p)\times$ SO$(q)$, for any arbitrary constant $c$.}.
Using the parameterization (\ref{paramer}), the explicit form of this metric is
\begin{equation}
	\begin{array}{rcl}
	ds^2/R^2 &=& \cosh(2 \psi) d \psi^2 + \cosh^2 \psi [d \theta_1^2 +\ldots + \sin^2 \theta_1 \ldots \sin^2 \theta_{p-2}\, d \theta_{p-1}^2] \\[3mm]
	&+& \sinh^2 \psi [d \phi_1^2 + \ldots + \sin^2 \phi_{1} \ldots \sin^2 \phi_{q-2}\, d \phi_{q-1}^2].
	\end{array}
\end{equation}
As we will see later, this is going to be the metric of the internal space in the dimensional reduction from M-theory to 4-dimensional supergravity where the 4-dimensional scalar fields are evaluated at the origin of the scalar $\sigma$-model.
Expectation values of the 70 scalars of the 4-dimensional theory correspond to deformations of this metric (like squashings of the spheres, rescalings of their size and different fibrations).

Before proceeding, it is also useful to introduce
\begin{equation}
K_{mn}{}^{AB} \equiv R^{-1}\,\stackrel{\circ}g_{L\,mp}\, \stackrel{\circ}{\nabla}_n K^{p AB},	
\end{equation}
which represent the covariant derivatives of the SO($p,q$) Killing vectors with respect to the non-euclidean metric.
These tensors obviously satisfy $K_{mn}{}^{AB} = -K_{nm}{}^{AB}$ and will become crucial in the construction of the 3-form ansatz.
It is also useful to note that such Killing vectors satisfy the orthogonality relations
\begin{equation}
 K_{mn}{}^{AB}\, K^p{}_{AB}=0\,.	\label{ortho}
\end{equation}


\subsection{Clifford properties and the new ansatz} 
\label{sub:clifford_properties_and_the_new_ansatz}

The generalized vielbeins (\ref{emu})--(\ref{emnu}) satisfy some important relations, which are necessary to ensure the consistency of the reduction process in the case of the SO(8) gauging and which we now show to hold also for more general SO($p,q$) gaugings.
The first relation is the so-called `Clifford property':
\begin{equation}
	e^{(m}{}_{ik}e^{n)\,kj}=\frac18 \delta_i{}^j e^m{}_{kl}e^{n\,kl}.\label{CP}
\end{equation}
In order to prove it we follow and generalize the steps in \cite{deWit:1986iy,deWit:2013ija}, where $(\ref{CP})$ is proved by showing that the left hand side vanishes when contracted with an arbitrary traceless matrix $\Lambda^i{}_j$.
This is equivalent to proving that $e^{(m}_{kl} e^{n)ij}$ vanishes when contracted with an arbitrary anti-hermitean traceless $\mathfrak{su}(8)$ matrix $\Lambda_{ij}{}^{kl} = \Lambda_{[i}{}^{[k}\delta_{j]}{}^{l]}$.
Inserting (\ref{emu})--(\ref{emd}) into (\ref{CP}) and using the symmetry in $m$, $n$, which is equivalent to the exchange $IJ \leftrightarrow KL$ in the expressions below, we get that $e^{(m}_{ij} e^{n)ij}$ is proportional to
\begin{equation}
	\tilde K^m{}_{IJ} \tilde K^n{}_{KL}  \Lambda^{ij}{}_{kl}\left[ (u_{ij}{}^{IJ} + v_{ijIJ})(u^{kl}{}_{KL}+v^{klKL}) + (IJ \leftrightarrow KL)\right],
\end{equation}
where we introduced
\begin{equation}
	\tilde K^m{}_{IJ} = -\frac12\, (\Gamma^{IJ})_{AB} K^m{}_{AB}, \qquad \tilde K^m{}^{IJ} = -\frac12\, (\Gamma^{IJ})_{AB} K^m{}^{AB}.
\end{equation}
This expression in square brackets multiplied by the $\Lambda$ matrix contains two types of terms.
The first one (and its conjugate) corresponds to a generic element of the $\mathfrak{su}(8)$ algebra:
\begin{equation}\label{su8rel}
	u^{kl}{}_{IJ} \Lambda_{kl}{}^{ij} u_{ij}{}^{KL} - v_{ijIJ} \Lambda^{ij}{}_{kl} v^{klKL} = \delta^{[K}_{[I} X^{L]}{}_{J]},
\end{equation}
with $X^\dagger = -X$ and $X^I{}_I = 0$, as follows from the matrix multiplication $U_{SU_8} \Lambda U_{SU_8}^\dagger = X$.
The second one (and its conjugate) is an antisymmetric tensor, which is the imaginary part of a ``complex selfdual'' tensor, which is antisefdual:
\begin{equation}
	u_{ij}{}^{IJ} \Lambda^{ij}{}_{kl} v^{kl KL} +u_{ij}{}^{KL} \Lambda^{ij}{}_{kl} v^{kl IJ}.
\end{equation}
Thanks to these relations we can see that $e^{(m}{}_{kj}e^{n)\,ij}\Lambda_{i}{}^{k}$ vanishes.
In fact, using (\ref{su8rel}), we obtain terms proportional to 
\begin{equation}
		\tilde K^{(m}{}_{IJ}\tilde K^{n)}{}_{IL} =\cosh(2\psi) \,g_E^{mn}\delta_{JL},\label{KKg}
\end{equation}
which vanish when contracted with $X^J{}_L$ because of the tracelessness of $X$.
The remaining terms are contracted with
\begin{equation}
	\tilde K^{(m}{}_{[IJ}\tilde K^{n)}{}_{KL]}, \label{KKSD}
\end{equation}
which is selfdual and therefore vanishes when contracted with an anti-selfdual tensor.
In this small proof, we chose $\Lambda$ antihermitean, but the result holds also for arbitrary hermitean matrices simply by repeating the same steps multiplying $\Lambda$ by $i$. 

The second important property satisfied by the generalized vielbeins, which we will need in the following, is that  
\begin{equation}
	e_p^{ij} \, e_{mn\,ij}= e_{[p}^{ij}\, e_{mn]\,ij},
\end{equation}
so that one establishes the antisymmetry of the 3-form tensor.
This expression is obviously antisymmetric in the last two indices and therefore we only need to prove antisymmetry in $[np]$ in the expression $\Delta^{-1} g^{nr} e^{p\,ij} \, e_{mr\,ij}$.
This expression gives terms proportional to
\begin{equation}
	\tilde K^n_{IJ} \tilde K^r_{KL} \tilde K_{mr}^{MN} \tilde K^{p}_{PQ} (u_{ij}{}^{IJ} + v_{ijIJ})(u^{ij}{}_{KL} + v^{ijKL})(u_{kl}{}^{MN}-v_{klMN})(u^{kl}{}_{PQ}+v^{klPQ}).
\end{equation}
The proof follows exactly the same argument as the one presented in \cite{deWit:2013ija}, from eq.~(5.17) onwards, by noting that
\begin{equation}
	\tilde K_{mr}{}^{MN}\tilde K^{r}{}_{PQ}=4g^{E}_{mr}\delta_{M[P} \tilde K^{r}{}_{Q]N}+ \tilde K_{mr}{}^{KL}
	\delta^{KL}_{[MN}\tilde K^{r}{}_{PQ]}, \label{so8Id}
\end{equation}
that $\tilde K_{mn}^{PQ}\delta^{PQ}_{[IJ}\tilde K^{n}{}_{KL]}$ is self-dual and the fact that our expressions match those in \cite{deWit:2013ija}, if we replace $K^{n\,IJ} \rightarrow\tilde K^n{}_{IJ}$ and $K_{mn}{}^{IJ} \rightarrow\tilde K_{mn}{}^{IJ}$. 

Now that we established that the new generalized vielbeins (\ref{emu})--(\ref{emnu}) satisfy the same properties as those that ensured the consistency of the reduction ansatz on $S^7$, we proceed by explicitly writing down the ansatz for the 11-dimensional metric and 3-form.
As in \cite{deWit:2013ija}, they follow by appropriate contractions of the generalized vielbeins.
In detail:
\begin{equation}
	\Delta^{-1} g^{mn} =\frac18\, e^{m\,ij}e^n_{ij} = K^m_{AB}(y) \, K^n_{CD}(y) \,{\cal V}^{AB\,ij}(x)\, {\cal V}^{CD}{}_{ij}(x),
\end{equation}
where
\begin{equation}
	\Delta(x,y)=\sqrt{\frac{det \left(g_{mn}(x,y)\right)}{\left|det \left(\stackrel{\circ}g_{L\,mn}(y)\right)\right|}}, 
\end{equation}
and
\begin{equation}
	A_{mnp} = -\frac18 \, \Delta\, g_{pq}\, e_{mn\,ij} e^{q\,ij} =  \frac{1}{\sqrt2} \, \Delta(x,y) \, g_{pq}(x,y) \, K_{mn}^{AB}(y)\,  K^q_{CD}(y)\,{\cal V}_{AB\,ij}(x)\,{\cal V}^{CD\,ij}(x).
\end{equation}
When we replace the explicit expressions for the coset representatives, we end up with explicit expressions in terms of the $u$ and $v$ matrices:
\begin{equation}
	\Delta^{-1} g^{mn} =\frac{1}{32}\, K^m{}_{AB} \, K^n{}_{CD} \,(\Gamma^{IJ})_{AB} (\Gamma^{KL})_{CD}\, (u^{ij}{}_{IJ}+v^{ijIJ})\, (u_{ij}{}^{KL}+v_{ijKL}), \label{metricansatz}
\end{equation}
\begin{equation}
	A_{mnp} = -\frac{i}{32\sqrt2} \, \Delta\, g_{pq}\, K_{mn}{}^{AB}\,  K^q{}_{CD}\,(\Gamma^{IJ})_{AB} (\Gamma^{KL})_{CD}\, (u^{ij}{}_{IJ}-v^{ijIJ})\,(u_{ij}{}^{KL}+v_{ijKL})\,. \label{3formansatz}
\end{equation}



\section{The maximally symmetric vacua of SO(5,3) and SO(4,4) gaugings} 
\label{sec:the_maximally_symmetric_vacua_of_so_5_3_and_so_4_4_gaugings}

We now perform a first test of our ansatz by checking that it reproduces the known M-theory solutions that lead to the maximally symmetric vacua of the SO(5,3) and SO(4,4) gauged supergravity models in 4 dimensions.
The 11-dimensional backgrounds we recover are hyperbolic geometries with trivial 3-form potentials.
This implies that at this stage we can only check the metric ansatz, but we will see in the next section that we can also test the 3-form ansatz by deforming one of these geometries to describe other vacua with a smaller residual symmetry.

\subsection{SO(4,4)} 
\label{sub:so_4_4}

The standard SO(4,4) gauging has a critical point preserving SO(4) $\times$ SO(4) \cite{Hull:1984rt}. 
In our parametrization, this critical point appears at the origin of the moduli space:
\begin{equation}
	u_{ij}{}^{kl} = \delta^{kl}_{ij}, \qquad v_{ijkl} = 0.
\end{equation}
Plugging these conditions into (\ref{3formansatz}) we obtain that $A_{mnp}=0$ because of the orthogonality condition (\ref{ortho}).
The internal metric can be obtained using (\ref{metricansatz}), which in this case, thanks to (\ref{KKg}), reduces to
\begin{equation}
	\Delta^{-1}\, g^{mn}(x,y)= \cosh(2\,\psi)\, g_E^{mn}(y),	
\end{equation}
where $g_E^{mn}$ is again the inverse of the Euclidean hyperbolic metric (\ref{euc}). 
By computing the determinant of this expression we can fix $\Delta$ and conclude that
\begin{eqnarray}
		g_{mn}(x,y)&=& \left(\cosh(2\,\psi)\right)^{-\frac13}\, g^E_{mn}(y),\\
		g_{\mu\nu}(x,y)&=& \left(\cosh(2\,\psi)\right)^{\frac23}\, \tilde g_{\mu\nu}(x),
\end{eqnarray}
where $\tilde g$ is the metric of 4-dimensional de Sitter spacetime.
This is exactly the solution found in \cite{Hull:1988jw}.


\subsection{SO(5,3)} 
\label{sub:so_5_3}

Also the SO(5,3) gauging has a critical point that fully preserve the maximal compact subgroup SO(5) $\times$ SO(3) \cite{Hull:1984ea}.
In this case, however, the critical point is found at a specific point in the moduli space that does not coincide with the origin.
If we call $s$ the scalar vacuum expectation value parametrizing the SO(5) $\times$ SO(3) invariant points of the theory, we find the following coset representative in the SU(8) basis 
\begin{eqnarray}
U_{SU_8}(s)=\exp \left(\begin{matrix}
0 & -\frac12s  \tilde X_{IJKL}     \cr
-\frac12 s \tilde X^{IJKL}   & 0
\end{matrix}\right), 
\end{eqnarray}
where
\begin{equation}
	\tilde X_{IJ KL} \equiv -\frac18 X_{AB} \Gamma^{AB}_{IJKL}\,, 
\end{equation}
and
\begin{equation}
	X_{AB}=\rm{diag}\left(-1,-1,-1,-1,-1,\frac53,\frac53,\frac53\right).
\end{equation}
The critical point is located at $s = s^* \equiv -\frac38 \log 3$.
In the SL(8,${\mathbb R}$) basis of $E_7$, the coset representative can be parametrized by a matrix $S$ in the fundamental representation of $SL(8)$ as the exponentiation of the $X$ generator above
\begin{equation}
 S= e^{-\frac12 s\,X},	
\end{equation}
which implies that the metric is 
\begin{equation}
	\begin{array}{rcl}
		\Delta^{-1}\, g^{mn}(x,y)&=&\frac12 K^{m}{}_{AB}(y)K^{n}{}_{CD}(y)S_{[\underline{a}}{}^{[A}S_{\underline{b}]}{}^{B]} 
		S_{[\underline{a}}{}^{[C}S_{\underline{b}]}{}^{D]}\\[3mm]
		 &=& \frac12 K^{m}{}_{AB}(y)K^{n}{}_{CD}(y)M^{AC} M^{BD},
 	\end{array}
\end{equation}
with
\begin{equation}
	M^{AB}= S_{\underline{a}}{}^A S_{\underline{a}}{}^B=3^{-\frac38}\, \rm{diag}\left(1,1,1,1,1,3,3,3\right).
\end{equation}
and we use the underline notation $\underline{a}$ to denote the local indices of the scalars.
Note that for any value of $s$ we have that 
\begin{equation}
	\left(u^{ij}{}_{IJ}-v^{ijIJ}\right)\left(u_{ij}{}^{KL}+v_{ijKL}\right)=\delta_{IJ}^{KL},
\end{equation}
which implies once again that the internal three form reduces to terms proportional to $K_{mn}{}^{AB} K^{p}{}_{AB}$, which vanish. 

Using the explicit form of the Killing vectors, we obtain the explicit form of the metric in terms of the coordinates
\begin{equation}
	\Delta^{-1}\, g^{mn}(x,y)= 3^{-\frac34} \left(-1+2\cosh(2\,\psi)\right)\; g_{E,\frac13}^{mn}(y),
\end{equation}
where $g_{E,c^2}^{mn}$ is the inverse of the squashed Euclidean hyperbolic metric (see footnote 2). 
After computing $\Delta$, we obtain the final result
\begin{eqnarray}
		g_{mn}(x,y)&=& 3^{\frac12} \left(-1+2\cosh(2\,\psi)\right)^{-\frac13}~ g^{E,\frac13}_{mn}(y),\\[2mm]
		g_{\mu\nu}(x,y)&=& 3^{-\frac14}\left(-1+2\cosh(2\,\psi)\right)^{\frac23}~ \tilde g_{\mu\nu}(x),
\end{eqnarray}	
which coincides with the solution of \cite{Hull:1988jw}. 



\section{Uplift of the SO(3) $\times$ SO(3) critical point} 
\label{sec:uplift_of_the_so_3_times_so_3_critical_point}

In this section we provide the main result of this work. 
We test the non linear ansatz of the metric and 3-form to uplift a new critical point of the SO(4,4) supergravity found in \cite{Dall'Agata:2012sx}.
As we will see, this is the first example of a vacuum of a maximal gauged supergravity with non-compact gauge group whose uplift requires a non-vanishing internal 3-form in order to satisfy the equations of motion of 11-dimensional supergravity.
Since the 4-dimensional vacuum has a positive cosmological constant, we will have also in this case a non-compact internal manifold, but, as we will show explicitly, the metric is not a simple hyperboloid anymore.

Following \cite{Dall'Agata:2012sx}, the SO(3) $\times$ SO(3) invariant critical points of SO(4,4) gauged supergravity can be obtained by truncating the scalar manifold to the scalars corresponding to the following E$_{7(7)}$ generators in the SL(8,${\mathbb R}$) basis (\ref{e7 gen in sl8 basis}):
\begin{equation}
	\begin{array}{rcl}
	g_5 &=& t_1{}^1+t_2{}^2+t_3{}^3+t_5{}^5+t_6{}^6+t_7{}^7-3(t_4{}^4+t_8{}^8),  \\[2mm]
	g_6 &=& t^{1238}+t^{4567}.
	\end{array}
\end{equation}
By taking $g_5$ and $g_6$ normalized so that Tr($g_i g_i$) =1, the associated coset representative is
\begin{equation}\label{Uxtau}
	{\cal U}(x, \tau) = \exp\left(\frac{3}{\sqrt2}\,g_5\, \log x +\sqrt{6} \,g_6\, \log \tau\right).
\end{equation}
In this parameterization the allowed moduli space is spanned by $x>0$ and $\tau>0$.
The explicit form of the scalar potential is \cite{Dall'Agata:2012sx}
\begin{equation}
	V(x, \tau) = \displaystyle \frac1{8 \,\tau\, x^{3/2}} \left[6 x  (\tau -1)^2+(1+\tau)^2-3 x^2 \left(1	-6 \tau +\tau^2\right) \right]
\end{equation}
and has the maximally symmetric vacuum discussed in sec.~\ref{sub:so_4_4} at $x=\tau=1$, but also shows another vacuum at 
\begin{equation}
x =  1+ \frac{2}{\sqrt3}, \qquad \tau_{\pm} = \sqrt3 \pm \sqrt2\,,
\end{equation}
where $\tau_{\pm}$ are identified using parity.
From (\ref{Uxtau}), we can derive the coset representatives in the mixed basis (\ref{coset representatives mixed basis}), which leads to the generalized vielbeins entering in the ansatz for the inverse internal metric.
It is interesting to note that ${\cal V}^{ABij}{\cal V}^{CD}{}_{ij}$ is diagonal in the $AB,CD$ indices and the coefficients respect the SO(3) $\times$ SO(3) symmetry, so that, if we set $a=1,2,3$, $\hat{a} = 5,6,7$, we get:
\begin{equation}
	\begin{array}{rcl}
	{\cal V}^{ab\,ij}{\cal V}^{cd}{}_{ij} &=& \displaystyle	{\cal V}^{\hat{a}\hat{b}ij}{\cal V}^{{\hat{c}\hat{d}}}{}_{ij} = \delta^{ac} \delta^{bd}\, \frac12 \sqrt{3 + 2 \sqrt3}, \\[3mm]
	{\cal V}^{{a}4ij}{\cal V}^{{c}4}{}_{ij} &=& \displaystyle	{\cal V}^{{\hat a}8ij}{\cal V}^{{\hat c}8}{}_{ij} = \delta^{ac}\, \frac12 \sqrt{-3 + 2 \sqrt3}, \\[3mm]
	{\cal V}^{{a\hat b}ij}{\cal V}^{{c \hat d}}{}_{ij} &=&\displaystyle \delta^{ac} \delta^{bd}\, \frac12\sqrt{1 + \frac{2}{\sqrt3}}, \\[3mm]
	{\cal V}^{{a}8ij}{\cal V}^{c8}{}_{ij} &=& \displaystyle	{\cal V}^{{\hat a}4ij}{\cal V}^{{\hat c}4}{}_{ij} = \delta^{ac}\, \frac32\sqrt{-1+\frac{2}{\sqrt3}}, \\[3mm]
	{\cal V}^{48ij}{\cal V}^{48}{}_{ij} &=&\displaystyle \frac{1}{2(1+2/\sqrt3)^{3/2}}.
	\end{array}
\end{equation}
This simplifies the inversion, from which one gets the final form of the metric:
\begin{equation}
	ds_7^2 = \frac{\alpha^{-1}R^2}{\Delta(y)} \left[ \sum_{i,j=1}^3 h_{ij}(y) dy^i dy^j 
+ R_1^2(y) \left(d\theta_2^2 + \sin^2(\theta_2)\;	 d\theta_3^2 \right) + R_2^2(y) \left(d\phi_2^2 + \sin^2(\phi_2)\; d\phi_3^2\right) \right],\label{metric}
\end{equation}
where we chose the coordinates $y^i = \{\psi,\theta_1,\phi_1, \theta_2, \theta_3, \phi_2, \phi_3\}$.

The metric (\ref{metric}) is a deformation of the hyperboloid at $x=\tau =1$, which preserves an $S^2 \times S^2$ within the original $S^3 \times S^3$ and therefore has a residual isometry group equal to SO(3) $\times$ SO(3).
In fact, the warp factors $\Delta,\, R_1$ and $R_2$ depend only on $y^1,y^2$ and $y^3$ and so does also the 3-dimensional metric $h_{ij}$.
In terms of the embedding coordinates, this means that all these functions depend only on the combinations 
\begin{equation}
	\begin{array}{rclcrcl}
	(X^1)^2+(X^2)^2+(X^3)^2 & = & R^2 \,\cosh^2 \psi\,\sin^2 \theta_1 &  & (X^4)^2 & = & R^2 \,\cosh^2 \psi\,\cos^2 \theta_1 ,\\[2mm]
	(X^5)^2+(X^6)^2+(X^7)^2 & = & R^2 \,\sinh^2 \psi\,\sin^2 \phi_1 &  & (X^8)^2 & = & R^2 \,\sinh^2 \psi\,\cos^2 \phi_1 .
	\end{array}
\end{equation}
In detail, the expression of the warp factors rescaling the spheres is
\begin{eqnarray}
R_1^2&=&\frac{4(2\sqrt3-3) \sin^2\theta_1}{3(\sqrt3-1)+6 \sin^2\theta_1+ \tanh^2\psi[3\sqrt3(\sqrt3-1)-(6-4\sqrt3) \sin^2\phi_1] }\;,\\[2mm]
		R_2^2&=&\frac{4(2\sqrt3-3) \sin^2\phi_1}{3(\sqrt3-1)+6 \sin^2(\phi_1)+ \coth^2\psi[3\sqrt3(\sqrt3-1)-(6-4\sqrt3) \sin^2\theta_1] }\;,
\end{eqnarray}
and the overall warp factors are
\begin{equation}
		\Delta^{-9} = \alpha^7 \,{\rm{det}}\,h^{-1}\;\frac{\sin^4\theta_1}{R_1^4} \frac{\sin^4\phi_1}{R_2^4}\cosh^6\psi\; \sinh^6\psi
\end{equation}
and
\begin{equation}
	\alpha^{-1}=\frac34(\sqrt3-1)\,\left(1+\frac{2}{\sqrt3}\right)^{\frac32}.
\end{equation}
The 3-dimensional metric that describe the mixing between the deformed circles of the spheres and the non-compact direction $\psi$ is 
\begin{equation}
	h = (\rm{det}\, M)^{-\frac12}\, M
\end{equation}
where
\begin{equation}
		M= \left(\begin{matrix}
		A\,B - S_2^2\; S_3^2 & -(\Phi_3\, B+ S3^2\;\Phi_2)\; S_2 & -(\Phi_2\, A + S2^2\;\Phi_3)\; S_3 \cr
		-(\Phi_3\, B+ S_3^2\;\Phi_2)  S_2 & [\frac{3(2+\sqrt3)}{4}-\Phi_2 \Phi_3]\,B-S_3^2\;\Phi_2^2 &  \frac{3(2+\sqrt3)}{4}\; S_2\; S_3  \cr
		-(\Phi_2\, A + S_2^2\;\Phi_3)\;  S_3  & \frac{3(2+\sqrt3)}{4}\; S_2\; S_3  &   [\frac{3(2+\sqrt3)}{4}-\Phi_2 \Phi_3]\,A-S_2^2\;\Phi_3^2
		\end{matrix}\right),
\end{equation}
and
\begin{equation}
	\begin{array}{rclcrcl}
		S_2 &=& \frac12\,\tanh\psi\, \sin(2 \theta_1), & & S_3 &=& \frac12\, \coth \psi\, \sin (2 \phi_1), \\[3mm]
		\Phi_2 &=& \frac32 - \sin^2\theta_1\,, && \Psi_2 &=& \frac32 - \cos^2\theta_1\,, \\[3mm]
		\Phi_3 &=& \frac32 - \sin^2\phi_1\,,&& \Psi_3 &=& \frac32 - \cos^2\phi_1\,, \\[3mm]
		A &=& \frac{3+\sqrt3}{4} + \tanh^2\psi \left[\frac{3(2+\sqrt3)}{4}-\Psi_2\Phi_3\right]\;,&&
	B&=&\frac{3+\sqrt3}{4} + \coth^2\psi \left[\frac{3(2+\sqrt3)}{4}-\Psi_3\Phi_2\right]\,.
	\end{array}
\end{equation}

Now we move to the computation of the internal three form. 
Contrary to the cases considered in the previous sections, the scalar matrices in (\ref{FAmixframe}) lead, in addition to a term proportional to the identity matrix, to some off diagonal components generating a non trivial internal three form. 
Also in this case the structure of the tensor we obtain respect the SO(3) $\times$ SO(3) symmetry, and the only non-vanishing components contain products of the Killing vectors such that the $A,B,C,D$ indices contributing are only in SO(3) $\times$ SO(3) invariant combinations:
\begin{equation}
	\begin{array}{rcrl}
	A_{mnp}(x,y)&=&\gamma~ \Delta~ g_{pq}(x,y)& \left[\left(x\; K_{mn}{}^{8[1}(y)K^{|q|23] }(y)+\; K_{mn}{}^{[12}(y)K^{|q|3]8}(y)\right)\right.\\[5mm]
	&&-&\left.\left(x\; K_{mn}{}^{4[5}(y)K^{|q|67] }(y)+\; K_{mn}{}^{[56}(y)K^{|q|7]4}(y)\right)\right]\,,
	\end{array}
\end{equation}
and
\begin{equation}
	\gamma=-3\sqrt{-3+2 \sqrt3}.
\end{equation}

After replacing the Killing tensors of the hyperboloid we get only six non vanishing 3-form components $\left(A_0=\frac{\gamma\; R^3}{3 \alpha }\right)$
\begin{equation}
	A=A^{(1)}\wedge e^{4}\wedge e^{5}+A^{(2)}\wedge e^{6}\wedge e^{7},
\end{equation}
where
\begin{eqnarray}
	e^4 &=& R_1 \, d \theta_2,\\[2mm]
	e^5 &=& R_1 \, \sin \theta_2\, d \theta_3,\\[2mm]
	e^6 &=& R_2 \, d \phi_2,\\[2mm]
	e^7 &=& R_2 \, \sin \phi_2 \, d \phi_3,
\end{eqnarray}
\begin{equation}
	\begin{array}{rcl}
	A^{(1)} & = & -A_0 \left[\sinh^2(\psi) + x\cosh^2(\psi)\right] \sin(\theta_1)\cos(\phi_1) \,d \psi  \\[2mm]
	&-&A_0 \sinh(\psi)\cosh(\psi)\cos(\theta_1)\cos(\phi_1)\, d \theta_1 \\[2mm]
	&+& A_0\, x\; \sinh(\psi)\cosh(\psi)\sin(\theta_1)\sin(\phi_1)\, d \phi_1
	\end{array}
\end{equation}
and
\begin{equation}
	\begin{array}{rcl}
	A^{(2)} & = & -A_0 \left[\cosh^2(\psi) + x\sinh^2(\psi)\right] \cos(\theta_1)\sin(\phi_1)\,d \psi  \\[2mm]
	&+& A_0 x\; \sinh(\psi)\cosh(\psi)\sin(\theta_1)\sin(\phi_1) \, d \theta_1\\[2mm]
	&-& A_0 \sinh(\psi)\cosh(\psi)\cos(\theta_1)\cos(\phi_1) \, d \phi_1.
	\end{array}
\end{equation}
As a result, the only non vanishing field strength components are 
\begin{eqnarray}
	\begin{array}{rcl}
	F_{ij45}&=&-F_{ij54}=4\partial_{[i}A_{j45]}=2 \partial_{[i}A_{j]54}=V_{ij}\;  R_1^2 \sin(\theta_2)\;,\\[3mm]    
	F_{ij67}&=&-F_{ij76}=4\partial_{[i}A_{j67]}=2 \partial_{[i}A_{j]67}=W_{ij}\;  R_2^2 \sin(\phi_2)\;,
	\end{array}
	\label{fieldstr}
\end{eqnarray}
where $V$ and $W$ are functions is $\psi,\theta_1,\phi_1$ and so they are again explicitly invariant under $SO(3)\times SO(3)$. 
The non-vanishing components of the two forms $V$ and $W$ are 
\begin{eqnarray}
	V_{12}&=& A_0 \cos(\theta_1)\cos(\phi_1) \times \nonumber\\[2mm]
	&&\left(2(2+\sqrt3)\cosh^2(\psi)-\frac{\sqrt3+3}{2}R_1^2\,\frac{(1-\sqrt3)-4(1+\frac{1}{\sqrt3})\,\cosh^2(\psi)\,\sin^2(\theta_1)}{(1-\sqrt3)\; \sin^2(\theta_1)}\right)\,,\nonumber \\[2mm]
	V_{13}&=& A_0 \sin(\theta_1)\sin(\phi_1)\tanh^2(\psi)\times\nonumber\\[2mm]
	&&\left(\frac{2}{\sqrt3}\cosh^2(\psi)+\frac{\sqrt3+3}{2}R_1^2\,\frac{(\frac53+\sqrt3)+\frac43(\sqrt3-1)\sinh^2(\psi)\cos^2(\phi_1)}{(1-\sqrt3)\; \sin^2(\theta_1)}\right)\,,\label{Vij}\\[2mm]
	V_{23}&=& A_0 \cos(\theta_1)\sin(\phi_1)\sinh(\psi)\cosh(\psi)\times\nonumber\\[2mm]
	&&\left(2(1+\sqrt3)-\frac{\sqrt3+3}{2}R_1^2\,\frac{-\left(2+\frac{4}{\sqrt3}\right)\,\sin^2(\theta_1)+\left(2-\frac{4}{\sqrt3}\right)\tanh^2(\psi)\cos^2(\phi_1)}{(1-\sqrt3)\; \sin^2(\theta_1)}\right)\,,\nonumber
\end{eqnarray}
and 
\begin{eqnarray}
	W_{13}&=& A_0 \cos(\theta_1)\cos(\phi_1) \times \nonumber \\[2mm]
	&& \left(2(2+\sqrt3)\sinh^2(\psi)-\frac{\sqrt3+3}{2}R_2^2\,\frac{(\sqrt3-1)-\frac43(3+\sqrt3)\,\,\sinh^2(\psi)\,\sin^2(\phi_1)}{(1-\sqrt3)\; \sin^2(\phi_1)}\right)\,,\nonumber\\[2mm]
	W_{12}&=& A_0 \sin(\theta_1)\sin(\phi_1)\coth^2(\psi)\times\nonumber\\[2mm]
	&&\left(\frac{2}{\sqrt3}\sinh^2(\psi)-\frac{\sqrt3+3}{2}R_2^2\,\frac{(\frac53+\sqrt3)+\frac43(1-\sqrt3)\cosh^2(\psi)\cos^2(\theta_1)}{(1-\sqrt3)\; \sin^2(\phi_1)}\right)\,,\label{Wij}\\[2mm]
	W_{23}&=& A_0 \sin(\theta_1)\cos(\phi_1)\sinh(\psi)\cosh(\psi)\times\nonumber\\[2mm]
	&&\left(-2(1+\sqrt3)+\frac{\sqrt3+3}{2}R_2^2\,\frac{-\left(2+\frac{4}{\sqrt3}\right)\,\sin^2(\phi_1)+\left(2-\frac{4}{\sqrt3}\right)\coth^2(\psi)\cos^2(\theta_1)}{(1-\sqrt3)\; \sin^2(\phi_1)}\right)\,.\nonumber
\end{eqnarray}

\subsection{Field equations}

Now that we have described in detail our ansatz for the metric and 3-form we put it at test against the equations of motion of 11-dimensional supergravity.
The field equations we will test are 
\begin{eqnarray}
	\!\!\!\!\!\!	D_M F^{MNPQ}&=&\!\!\!\frac{1}{\sqrt{|g|}}\partial_M(\sqrt{|g|}\, F^{MNPQ})=-\frac{\sqrt{2}}{2 (4!)^2}\epsilon^{EFGHIJKLNPQ}F_{EFGH}F_{IJKL} \,,\label{MaxEquation} \\[3mm]
	R_{MN}&=&-\frac16\left(F_{MPQR} F_N{}^{PQR}-\frac{1}{12} F_{PQRS} F^{PQRS} g_{MN}\right)\,.\label{EinEquation}
\end{eqnarray} 
We will use the following splitting for the 7 dimensional indices:
$y^i = \{\psi,\theta_1,\phi_1\}$, $y^a =\{\theta_2,\theta_3\}$, $y^{\hat a} = \{\phi_2,\phi_3\}$.
The only non vanishing terms in the field strengths are, up to permutations, $F_{\mu\nu\rho\sigma}$, $F_{ijab}$ and $F_{ij\hat a\hat b}$. 
The external components of the field strength are fixed by the Freund-Rubin solution (\ref{4DF}), which is clearly a solution of (\ref{MaxEquation}). 	

After inserting the ansatz (\ref{4Dmetric}), (\ref{metric}) and (\ref{fieldstr}), the only non trivial constraints imposed by Maxwell's equation (\ref{MaxEquation}) come from $D_i F^{ijab}$ and $D_i F^{ij\hat a\hat b}$. 
They lead to
\begin{equation}
	\begin{array}{rcl}
	\displaystyle \partial_i \left(\Delta^{-2}\, R_2^2\, \sqrt{det(g_{ij})}\, V^{ij}\right)&=&\displaystyle-\frac{\sqrt2}{2}\, R_2^2\, f_{FR}\;\varepsilon^{jkl} W_{kl},\\[2mm]
	\displaystyle \partial_i \left(\Delta^{-2}\,R_1^2\, \sqrt{det(g_{ij})}\, W^{ij}\right)&=&\displaystyle-\frac{\sqrt2}{2}\, R_1^2\, f_{FR}\;\varepsilon^{jkl} V_{kl},
	\end{array}\label{MaxEq}
\end{equation}
where the internal $i,j,k$ indices are raised with $g^{ij}=\alpha\,R^{-2}\, \Delta h^{ij}$ and $\varepsilon_{ijk}$ is the 3 dimensional Levi Civita symbol. 

The non-trivial terms in Einstein's equations can be computed by noting that 
\begin{eqnarray}
	\begin{array}{rcl}
	F_{\mu PQR}F_{\nu}{}^{PQR}&=&-3!f_{FR}^2\;\Delta^4\, g_{\mu\nu}(x,y)\,,\\[3mm]
	F_{a PQR}F_{b}{}^{PQR}&=&3!\left(\alpha\,R^{-2}\, \Delta\right)^2 \frac12 V^2 \;g_{ab}\,,\\[3mm]
	F_{\hat{a}PQR} F_{\hat{b}}{}^{PQR}&=&3!\left(\alpha\,R^{-2}\, \Delta\right)^2 \frac12 W^2\; g_{\hat{a}\hat{b}}\,,\\[3mm]
	F_{i PQR}F_{j}{}^{PQR}&=&3!\left(\alpha\,R^{-2}\, \Delta\right)^2 \left(V_{i}{}^{k} V_{jk}+W_{i}{}^{k} W_{jk}\right)\,,\\[3mm]
	F_{MNPQ}F^{MNPQ}&=&4!\left(-f_{FR}^2\,\Delta^4 + \frac12  \left(\alpha\,R^{-2}\, \Delta\right)^2 \left(V^2+W^2\right)\right)\,,
	\end{array}\label{F2}
\end{eqnarray}
where $V^2=V_{ij}V^{ij},W^2=W_{ij}W^{ij}$.
Plugging (\ref{F2}) into (\ref{EinEquation}) and using (\ref{metric}) for the metric, we obtain the following set of constraints
\begin{equation}\label{EinEq}
	\begin{array}{rcl}
	R_{\mu\nu}&=&\displaystyle\frac{1}{6}\left(4\;f_{FR}^2\, \Delta^4+\left(\alpha\,R^{-2}\, \Delta\right)^{2} \left(V^2+W^2\right)\right)g_{\mu\nu}\,,\\[4mm]
	R_{ij}&=&\displaystyle-\frac{1}{6}\left[\left(2\;f_{FR}^2\, \Delta^4-\left(\alpha\,R^{-2}\, \Delta\right)^{2}\left(V^2+W^2\right)\right)g_{ij} +
	3!\left(\alpha\,R^{-2}\, \Delta\right)^{2}\left(V_{i}{}^{k} V_{jk}+W_{i}{}^{k} W_{jk}\right)\right]\,,\\[4mm]
	R_{ab}&=&\displaystyle-\frac{1}{6}\left(2\; f_{FR}^2\, \Delta^4+\left(\alpha\,R^{-2}\, \Delta\right)^{2} \left(2 V^2-W^2\right)\right)g_{ab}\,,\\[4mm]
	R_{\hat a\hat b}&=&\displaystyle-\frac{1}{6}\left(2\; f_{FR}^2\, \Delta^4+\left(\alpha\,R^{-2}\, \Delta\right)^{2} \left(2W^2-V^2\right)\right) g_{\hat a\hat b}\,.
	\end{array}
\end{equation}
It is interesting to note that these components are not linearly independent. 
Indeed it holds that\footnote{It is worth noticing that these equations are a generalization of the standard equation
$$
	\frac 54 R_{\mu}{}^{\mu} + R_{m}{}^{m}= \Delta^4 f_{FR}^2, 
$$
coming from the conditions that the 4 dimensional space time is de Sitter and the only nonvanishing components of the field strength are $F_{\mu\nu\rho\sigma}$ and $F_{mnpq}$. See, for instance, eq.~(3.7) in \cite{deWit:1984nz}.}
\begin{equation}
	\begin{array}{rcl}
	\displaystyle\frac 34 R_{\mu}{}^{\mu} + R_{i}{}^{i}&=&\Delta^4 f_{FR}^2\,,\\[3mm]
	\displaystyle R_{\mu}{}^{\mu} +2(R_{a}{}^{a} + R_{\hat a}{}^{\hat a})&=&0.
	\end{array}\label{constrR}
\end{equation}
On the other hand, the non linear ansatz for the metric leads to the following non vanishing components for the Ricci tensor
\begin{equation}
	\begin{array}{rcl}
	R_{\mu\nu}&=&\tilde R_{\mu\nu}-\frac12 g^{ij}\left(\frac{\nabla_i\nabla_j \Delta^{-1}}{\Delta^{-1}}
	+\frac{\nabla_i\Delta^{-1}}{\Delta^{-1}} \left(\frac{\nabla_j\Delta^{-1}}{\Delta^{-1}}+\frac{\nabla_j( R_1^2 \Delta^{-1})}{R_1^2\Delta^{-1}}+\frac{\nabla_j( R_2^2 \Delta^{-1})}{R_2^2\Delta^{-1}}\right)\right) g_{\mu\nu}\,,\\[4mm]
	R_{ab}&=&\tilde R_{ab}-\frac{1}{2} g^{ij}\left(
	\frac{\nabla_i\nabla_j\left(R_1^2 \Delta^{-1}\right)}{R_1^2 \Delta^{-1}}
	+\frac{\nabla_i\left(R_1^2 \Delta^{-1}\right)}{R_1^2 \Delta^{-1}}
	\left(2\frac{\nabla_j\Delta^{-1}}{\Delta^{-1}}+\frac{\nabla_j( R_2^2 \Delta^{-1})}{R_2^2\Delta^{-1}}\right)\right)g_{ab}\,,\\[4mm]
	R_{\hat a\hat b}&=& \tilde R_{\hat a\hat b}-\frac{1}{2} g^{ij}\left(
	\frac{\nabla_i\nabla_j\left(R_2^2 \Delta^{-1}\right)}{R_2^2 \Delta^{-1}}
	+\frac{\nabla_i\left(R_2^2 \Delta^{-1}\right)}{R_2^2 \Delta^{-1}}
	\left(2\frac{\nabla_j\Delta^{-1}}{\Delta^{-1}}+\frac{\nabla_j (R_1^2 \Delta^{-1})}{R_1^2 \Delta^{-1}}\right)\right)g_{ab}\,,\\[4mm]
	R_{ij}&=&\tilde R_{ij}-\left(2\frac{\nabla_i\nabla_j \Delta^{-1}}{\Delta^{-1}}+\frac{\nabla_i\nabla_j \left(R_1^2 \Delta^{-1}\right)}{R_1^2 \Delta^{-1}}+
	\frac{\nabla_i\nabla_j \left(R_2^2 \Delta^{-1}\right)}{R_2^2 \Delta^{-1}}\right) \,,\\[4mm]
	&& +\frac12\left(2\frac{\nabla_i\Delta^{-1}}{\Delta^{-1}}\frac{\nabla_j\Delta^{-1}}{\Delta^{-1}}+\frac{\nabla_i\left(R_1^2 \Delta^{-1}\right)}{R_1^2 \Delta^{-1}}
	\frac{\nabla_j\left(R_1^2 \Delta^{-1}\right)}{R_1^2 \Delta^{-1}}+\frac{\nabla_i\left(R_2^2 \Delta^{-1}\right)}{R_2^2 \Delta^{-1}}
	\frac{\nabla_j\left(R_2^2 \Delta^{-1}\right)}{R_2^2 \Delta^{-1}}\right)\,	
	\end{array}
	\label{RicciExpansion}
\end{equation}
where $\tilde R_{\mu\nu},\tilde R_{ab},\tilde R_{\hat a\hat b},\tilde R_{ij}$ are the Ricci tensors computed with $\tilde g_{\mu\nu}, \tilde g_{ab}=\alpha\,R^{-2}\, \Delta\, g_{ab},\tilde g_{\hat a\hat b}=\alpha\,R^{-2}\, \Delta\, g_{\hat a\hat b}$ and $g_{ij}$ respectively.

The metric $\tilde g$ describes de Sitter spacetime and therefore $\tilde  R_{\mu\nu}=3\,R_4^{-2}\, \tilde g_{\mu\nu}$, with $R_4$ the de Sitter radius. 
Inserting this into (\ref{constrR}) we could then solve for $R_4$ and $f_{FR}$ finding
\begin{eqnarray}
R_4^{2}=\frac32 \frac{g^2}{V_{c}}R^2\,,\label{R4}
\end{eqnarray}
and
\begin{eqnarray}
f_{FR}=\pm\frac{1}{g^2\sqrt2}V_{c}R^{-1}\,,\label{fFR}
\end{eqnarray}
where $g$ is the coupling constant of the 4-dimensional gauged supergravity theory and $V_{c}$ is the value of the potential at the critical point
\begin{eqnarray}
V_{c}= 2 \, g^2\,\sqrt{-9 + 6 \sqrt3} \,.
\end{eqnarray}
These expressions are a highly non trivial check of the uplift ansatz and put it on very solid ground. 
Equation (\ref{R4}) is the generalization of the analogous one found for compact solutions \cite{Godazgar:2014eza}. 
It differs from our expression by a global sign (critical points in the compact case always have $V_c<0$), to guarantee the positivity of the proportionality factor between $R_4^2$ and $R^2$. 
Equation (\ref{fFR}) exactly agrees with the conjectured expression for the compact solution \cite{Godazgar:2014eza}. 
This is a very remarkable fact, because there is no proof of this formula, but all known uplift solutions satisfy it.

The complexity of the solution makes it difficult to provide a completely analytic proof, but it is not difficult to check with a computing program that the remaining 11-dimensional Maxwell and Einstein equations do not give any other constraint and are identically solved. 


\section{Hyperbloids as generalized twisted tori} 
\label{sec:hyperbloids_as_generalized_twisted_tori}

In this section we will show that the same ansatz for the metric and 3-form follows from some form of exceptional generalized geometry \cite{Lee:2014mla}.
This allows us to interpret such dimensional reductions as generalized Scherk--Schwarz reductions on non-trivial hyperbolic backgrounds. 
This interpretation was successfully displayed in the case of compactification on spheres \cite{Lee:2014mla}, and here we extend it to the case of the non compact hyperboloids.

\subsection{$S^7$ as a generalized twisted torus}

Before discussing our solutions, we review the construction of \cite{Lee:2014mla} for the seven sphere, in order to fix our notations and conventions.

The generalized vielbein ${\cal E}(x,y)$ encoding the degrees of freedom of the internal part of the metric and 3-form describing internal backgrounds with topology of 7 spheres can be written in the {\bf 21}+{\bf 7}+{\bf 21}'+{\bf 7}' decomposition as
\begin{eqnarray}
\left(\begin{matrix}
\Delta^{\frac12}\,e_{a}{}^{[m}\,e_{b}{}^{n]}              &                   0                                          &  
             0                                            &    \frac{\sqrt2}{2} \Delta^{\frac12}\,e_{a}{}^p\,e_{b}{}^q A_{mpq}   \cr
2\Delta^{\frac12}\,e_{\underline{a}}{}^p S_-^{mn}{}_p     &    \Delta^{-\frac12}\,e_{a}{}^m                              &  
\sqrt2\Delta^{-\frac12}\,e_{a}{}^p A_{mnp}               &    \Delta^{\frac12}\,e_{a}{}^p S_{mp}                       \cr
-\Delta^{\frac12}\,e^{a}{}_p\,e^{b}{}_q S^{mnpq}          &                   0                                          &   
\Delta^{-\frac12}\,e^{a}{}_{[m} \,e^{b}{}_{n]}            &    -\Delta^{\frac12}\,e^{a}{}_p\,e^{b}{}_q S_+^{pq}{}_m     \cr
             0                                            &                   0                                          & 
						 0                                            &    \Delta^{\frac12}\,e^{a}{}_m                              \cr
\end{matrix}\right),
\end{eqnarray}
with $e^{a}{}_m$ and $A_{mnp}$ representing the internal components of the vielbein and 3 form of the 11-dimensional supergravity, so that they have implicit dependence on all 11 coordinates. 
The other tensors appearing above are defined as
\begin{eqnarray}
S_{\pm}^{mn}{}_{p}&=&3\sqrt2\stackrel{\circ}\epsilon{}^{mn q_1\dots q_5}
\frac{1}{6!}\left(\tilde A_{p q_1\dots q_5}\pm 5\sqrt2  A_{p q_1 q_2}A_{q_3 q_4 q_5}\right),\\[2mm]
S_{mn}&=&-\frac{1}{5!}\stackrel{\circ}\epsilon{}^{q_1\dots q_7}A_{m q_1 q_2}\left(\tilde A_{n q_3\dots q_7}
-\frac{\sqrt2}{3} 5A_{n q_3 q_4}A_{q_5 q_6 q_7}\right),\\[2mm]
S^{mnpq}&=&\frac{\sqrt2}{3!\,2}\stackrel{\circ}\epsilon{}^{mnpq r_1 r_2 r_3} A_{r_1 r_2 r_3},
\end{eqnarray}
where $\tilde A$ is the dual six form, defined by
\begin{eqnarray}
\partial_{[M_1}\tilde A_{M_2\dots M_7]}=-\frac{1}{4! 7}\epsilon_{M_1\dots M_{11}}F^{M_8\dots M_{11}}-\frac{5\sqrt2}{2}A_{M_1 M_2 M_3}F_{M_4\dots M_7}
\end{eqnarray}
and, finally, $\stackrel{\circ}\epsilon_{q_1\dots q_7}$ and $\epsilon_{M_1\dots M_{11}}$ denote, respectively the volume form of the seven sphere and the one of the 11 dimensional manifold\footnote{Here $M_i$ denotes the 11 D curved indices splitting in $X^{M}=\left(x^\mu,y^m\right)$ and must not be confused with the SL(8) indices $MN$ in the generalized vielbein.}.
From the point of view of Exceptional Geometry, we chose the physical section depending on the 7 internal coordinates $y^m=X^{m\,8}$, so that the matrix block $\Delta^{\frac12}\,e_a{}^{[m}\,e_b{}^{n]}$ is identified with ${\cal E}_{ab}{}^{mn}$, $\Delta^{-\frac12}\,e_a{}^m $ with $2\, {\cal E}_{a8}{}^{m8}$ and so on.

The generalized vielbeins ${\cal E}(x,y)$ are coset representatives of $E_7/SU(8)$ and in the particular gauge chosen above they have the advantage of being triangular in the decomposition {\bf 7}+{\bf 21}'+{\bf 7}'+ {\bf 21}.
They also have a nice 11-dimensional geometric interpretation, but, as we will now see, this is not the best gauge choice if one is interested in discussing the Scherk--Schwarz reduction. 
To find a simpler gauge to compare with the results in \cite{deWit:1984nz,deWit:2013ija} and in our work, we use the non linear uplift ansatz for the metric, 3-form and 6-dual form, leading to the following decomposition
\begin{eqnarray}
{\cal E}_{\mathbb A}{}^{\mathbb M}(x,y)={\cal W}_{\mathbb A}{}^{\mathbb B}(y) 
U_{\mathbb B}{}^{\mathbb C}(x) E_{\mathbb C}{}^{\mathbb M}(y),
\end{eqnarray}
where 
\begin{eqnarray}
W_{\mathbb A}{}^{\mathbb B}=\frac{1}{16}\left(\begin{matrix}
\Gamma_{AB}^{\underline{I}\underline{J}}\hat\eta_{\underline{I}}{}^{K} 
      \hat\eta_{\underline{J}}{}^{L} \Gamma^{CD}_{K L}       &    0           \cr
  0   &   \Gamma^{AB}_{\underline{I}\underline{J}}\hat\eta^{\underline{I}}{}_{K} 
		\hat\eta^{\underline{J}}{}_{L} \Gamma_{CD}^{K L}  
\end{matrix}\right).
\end{eqnarray}
Here $\hat\eta_{\underline{I}}{}^{K}=(\hat\eta^{K})_{\underline{I}}$ denote the Killing spinors of the 7-sphere, $U$ denotes the scalar matrix in the $SL(8)$ frame and $E$ is given in terms of the Killing vectors, their covariant derivatives and the dual 6 form of $S^7$. 
Since $W(y)$ is actually an $SO(8)$ matrix, it can be reabsorbed by a gauge transformation and we conclude that despite the non linear mixing between internal and 4 dimensional coordinates (\ref{metricansatz}), (\ref{3formansatz}), the reduction has a linear decomposition in term of the generalized vielbeins and can be interpreted as a generalized Scherk--Schwarz reduction with an internal twist $E(y)$:
\begin{eqnarray}
\tilde{\cal E}_{\mathbb A}{}^{\mathbb M}(x,y)=U_{\mathbb B}{}^{\mathbb C}(x) E_{\mathbb C}{}^{\mathbb M}(y),
\end{eqnarray}
After dimensional reduction, the internal part of the local symmetries of 11--dimensional supergravity lead to the gauge symmetries of the effective theory and the corresponding embedding tensor is determined in terms of $E$ as a generalized version of the metric fluxes of standard twisted tori. 
In fact, the structure constants of the 4-dimensional supergravity theory appear in the generalized Lie derivatives (see equation (\ref{LieDerE})) of Exceptional Generalized Geometry (EGG) \cite{Lee:2014mla}.


\subsection{Scherk--Schwarz reductions on hyperboloids} 
\label{sub:scherk_schwarz_reductions_on_hyperboloids}

After reviewing the general idea of generalized Scherk--Schwarz reductions in EGG, we show that by applying this construction to the hyperboloids we can recover the non linear ansatz we provided for the metric and 3-form. 
In fact, both the metric and the 3-form are easily recovered from the generalized metric ${\cal H}={\cal E}^\dagger{\cal E}$, 
\begin{eqnarray}
\frac12\Delta^{-1}g^{mn}={\cal H}^{m8,\, n8}=
U_{\mathbb A}{}^{\mathbb B}(x) U_{\mathbb A}{}^{\mathbb C}(x) E_{\mathbb B}{}^{m8}(y)E_{\mathbb C}{}^{n8}(y),
\label{MA}
\end{eqnarray}
\begin{eqnarray}
\frac{\sqrt2}{2}\Delta^{-1}A_{mnq}g^{qp}={\cal H}_{mn}{}^{p8}=
U_{\mathbb A}{}^{\mathbb B}(x) U_{\mathbb A}{}^{\mathbb C}(x) E_{\mathbb B}{}_{mn}(y) E_{\mathbb C}{}^{p8}(y).
\label{FA}
\end{eqnarray}
All we need is to find the correct expression for $E$. We will do that below by extending the construction of \cite{Lee:2014mla} to the non-compact gauge groups SO($p,q$).

In the EGG formalism (see $e.g.$ \cite{Coimbra:2011ky,Coimbra:2012af}) the generalized vielbein discussed so far is alternatively represented as a direct sum on the generalized bundle
\begin{eqnarray}
V=v+w+\sigma+\tau~~ \in~~ TM\oplus \Lambda^2 T^*M\oplus\Lambda^5T^*M\oplus\left(T^*M\otimes \Lambda^7 T^*M\right).
\end{eqnarray}
The map between the two formalisms is given by
\begin{eqnarray}
E_{\mathbb A}{}^{m8}&=&-\frac12\, v_{\mathbb A}{}^m\,, \\[2mm]
E_{\mathbb A}{}^{mn}&=&\frac12\,\frac{1}{5!} \sigma^{\mathbb A}{}_{\; p_1\dots p_5}\stackrel{\circ}\epsilon{}^{p_1\dots p_5 mn}\,,\\[2mm]
E^{\mathbb A}{}_{m8}&=&\,-\frac12\,\frac{1}{7!}\,\tau^{\mathbb A}{}_{m,p_1\dots p_7} \stackrel{\circ}\epsilon{}^{p_1\dots p_7}\,,\\[2mm]
E^{\mathbb A}{}_{mn}&=&\frac12 w^{\mathbb A}{}_{mn}.
\end{eqnarray}

The generalized bundle is equipped with a generalized Lie derivative, 
\begin{equation}
	\begin{array}{rcl}
L_V V'&=&{\mathcal L}_v v' + \left({\mathcal L}_v w'-i_{v'}dw\right)+\left({\mathcal L}_v \sigma'-i_{v'}d\sigma-w'\wedge dw\right)\\[2mm]
&&+\left({\mathcal L}_v \tau'-j\sigma'\wedge dw-jw'\wedge d\sigma\right),\label{LieDer}
	\end{array}
\end{equation}
where the j-operation is defined as
\begin{eqnarray}
\left(j \alpha ^{(p+1)}\wedge \beta^{(d-p)}\right)_{m,n_1\dots n_d}=\frac{d!}{p!(d-p)!}\alpha_{m[n_1\dots n_p}\beta_{n_{p+1}\dots n_d]}.
\end{eqnarray}
Using this generalized Lie derivative we assume that the generalized Vielbein satisfies (\ref{LieDerE}), where $X_{\mathbb A\, \mathbb B}{}^{\mathbb C}$ gives the structure constants of the 4-dimensional gauge supergravity under examination, namely
\begin{equation}
	L_{E_{\mathbb A}} E_{\mathbb B} = X_{{\mathbb A}{\mathbb B}}{}^{\mathbb C} E_{\mathbb C}.
\end{equation}
For the $SO(p,q)$ gauge groups, we must therefore find a frame $E$ that solves 
\begin{eqnarray}
L_{E_{AB}}E_{CD}&=&R^{-1}\left(\eta_{CB}\,E_{AD}-\eta_{DB}\,E_{AC}
-\eta_{CA}\,E_{BD}+\eta_{DA}\,E_{BC}\right),\label{LieDerEdd} \\[2mm]
L_{E_{AB}}E^{CD}&=&R^{-1}\left(\eta_{AE}\, \delta_B^C\, E^{ED}
-\eta_{BE}\, \delta_A^C\, E^{ED}+\eta_{AE} \delta_B^D\, E^{CE}
-\eta_{BE} \delta_A^D\, E^{CE}\right), \quad\label{LieDerEdu} \\[2mm]
L_{E^{AB}}E_{CD}&=&0\,,\qquad L_{E^{AB}}E^{CD}=0\,.\label{LieDerEu}
\end{eqnarray}
In full analogy with the $S^7$ case, the solution is given by
\begin{eqnarray}
E_A=\left\{
\begin{array}{lcr}
E_{AB}=K_{AB}+S_{AB}+i_{K_{AB}}\zeta &&\\[3mm]
E^{AB}=P^{AB}+T^{AB}-j\zeta\wedge P^{AB},&&
\end{array}\label{vielgen}
\right.
\end{eqnarray}
with (note that $P^{AB}{}_{mn}=-K_{mn}{}^{AB}$)
\begin{eqnarray}
P^{AB}&=&dX^A\wedge dX^B, \nonumber\\[2mm]
S_{AB}&=&*P_{AB}=\frac{R^{-1}}{(d-2)!}\epsilon_{ABC_1\dots C_{d-1}}X^{C_1} dX^{C_2}\wedge\dots\wedge dX^{C_{d-1}}\\[2mm]
T^{AB}&=&R^{-1}\left(X^A dX^B-X^B dX^A\right)\otimes Vol_H,
\end{eqnarray}
where we set $\epsilon_{123\dots d+1}=1$, \emph{Vol}$_H$ denotes the volume form 
\begin{equation}
Vol_H=\frac{R^{-1}}{d!}~\epsilon_{C_1\dots C_{d+1}}X^{C_1} dX^{C_2}\wedge\dots\wedge dX^{C_{d+1}},
\end{equation}
and $\zeta$ its potential,
\begin{equation}
d\zeta=\frac{d-1}{R}Vol_H\,.\label{dualform}
\end{equation}
For us $d=7$ and therefore $\zeta$ is proportional to the dual 6 form $\zeta=-\sqrt2\,\tilde A$.
Note that the $E_A$ vielbeins defined in (\ref{vielgen}) are nowhere vanishing. 
While $T^{AB}$ vanishes at points where $X^A = X^B = 0$, the corresponding $P^{AB}$ either vanishes at points where $(X^A)^2 + (X^B)^2 = R^2$, for $A,B =1,\ldots,4$, or where $(X^A)^2 - (X^B)^2 = R^2$, for $A=1,\ldots,4$ and $B=5,\ldots,8$, or is never vanishing, for $A,B = 5,\ldots, 8$.
A similar argument applies to $K_{AB}$ and $S_{AB}$.
This means that also our euclidean hyperboloids are generalised parallelisable manifolds as defined in \cite{Lee:2014mla}.

As expected, the bracket of the Killing vectors reproduces the SO$(p,q)$ algebra
\begin{eqnarray}
\left[K_{AB},K_{CD}\right]=R^{-1}\left(\eta_{CB}\,K_{AD}-\eta_{DB}\,K_{AC}-\eta_{CA}\,K_{BD}+\eta_{DA}\,K_{BC}\right)\,.\label{KillLieDer}
\end{eqnarray}

In order to proceed with the proof, it is useful to note that
\begin{equation}
	\begin{array}{rcl}
{\mathcal L}_{K_{AB}}\; X^C=R^{-1}X^D\left(\eta_{DA}\delta_{B}^{C}-\eta_{DB}\delta_{A}^{C}\right)\,,\\[2mm]
{\mathcal L}_{K_{AB}}\; dX^C=R^{-1}dX^D\left(\eta_{DA}\delta_{B}^{C}-\eta_{DB}\delta_{A}^{C}\right).
	\end{array}\label{inetrmeqs}
\end{equation}
This implies that
\begin{eqnarray}
{\mathcal L}_{K_{AB}}\; P^{CD}=R^2\left(\eta^{EA}\, \delta_B^C\, P^{ED}
-\eta^{EB}\, \delta_A^C\, P^{ED}+\eta^{EA}\, \delta_B^D\, P^{CE}
-\eta^{EB}\, \delta_A^D\, P^{CE}\right)\label{wLieDer}
\end{eqnarray}
and using that $S_{AB}\wedge P^{AB}=d(d-1) Vol_H$ is invariant, one concludes that
\begin{eqnarray}
{\mathcal L}_{K_{AB}}\;	S_{CD}=R^{-1}\left(\eta_{CB}\,S_{AD}
-\eta_{DB}\,S_{AC}-\eta_{CA}\,S_{BD}
+\eta_{DA}\,S_{BC}\right)\label{sigmaLieDer}
\end{eqnarray}
and
\begin{equation}
	\begin{array}{rcl}
		{\mathcal L}_{K_{AB}}\;T^{CD}&=&{\mathcal L}_{K_{AB}}
\left[R^{-1}\left(X^C dX^D-X^D dX^C\right)\right]\otimes Vol_g\\[3mm]
	&=& R^{-1}\left(\eta^{AE}\,\delta_B^C\,T^{ED}-\eta^{BE}\,\delta_A^C\,T^{ED}
+\eta^{AE}\,\delta_B^D\,T^{CE}-\eta^{BE}\,\delta_A^D\,T^{CE}\right)\,.
	\end{array}
	\label{tauLieDer}
\end{equation}

We can finally prove (\ref{LieDerEdd}) by using (\ref{LieDer}), (\ref{KillLieDer}), (\ref{sigmaLieDer}) and 
\begin{equation}
	\begin{array}{rcl}
		{\mathcal L}_{K_{AB}}\left(i_{K_{CD}}\zeta\right)&-&i_{K_{CD}}d\left(S_{AB}
+i_{K_{AB}}\zeta\right)\;=\\[2mm]
	&=&{\mathcal L}_{K_{AB}}\left(i_{K_{CD}}\zeta\right) -i_{K_{CD}}\left({\mathcal L}_{K_{AB}} \zeta\right) 	-i_{K_{CD}}\left(dS_{AB}-i_{K_{AB}}d\zeta\right)\\[2mm]
	&=&i_{[K_{AB},K_{CD}]}\zeta\\[2mm]
	&=&R^{-1}\left(\eta^{CB}\,i_{K_{AD}}\zeta 
-\eta^{DB}\,i_{K_{AC}}\zeta-\eta^{CA}\,i_{K_{BD}}\zeta+\eta^{DA}\,i_{K_{BC}}\zeta\right),
	\end{array}
\end{equation}
where the last term in the second line vanishes because of (\ref{dualform}) and
\begin{eqnarray}
i_{K_{AB}}Vol_H=\frac{R}{d-1} dS_{AB}.\label{dsigma}
\end{eqnarray}
It is worth mentioning that this solution can also be obtained directly from the Extended Geometry point of view by following exactly the same steps of \cite{Baron:2014yua} for the compact gaugings.

The remaining generalized Lie derivative (\ref{LieDerEdu}) follows from (\ref{wLieDer}), (\ref{tauLieDer}), using also that 
\begin{eqnarray}
{\mathcal L}_{K_{AB}}\left(-j\zeta\wedge P^{CD}\right)
&-&j P^{CD}\wedge d\left(S_{AB}+i_{K_{AB}}\zeta\right) \\[3mm]
&=&{\mathcal L}_{K_{AB}}\left(-j\zeta\wedge P^{CD}\right)
-j P^{CD}\wedge \left({\mathcal L}_{K_{AB}}\zeta\right)\\[3mm]
&=&-j\zeta\wedge\left({\mathcal L}_{K_{AB}}P^{CD}\right).
\end{eqnarray}
Finally, (\ref{LieDerEu}) is a consequence of the fact that $P^{AB}$ is a closed two form.

After we proved that such generalized vielbeins satisfy the correct generalized Lie derivatives, we can use them as a starting point for the construction of an ansatz for the metric and 3-form.
By replacing the form of $E$ we introduced above into (\ref{MA}) and (\ref{FA}) we obtain
\begin{eqnarray}
\Delta^{-1}g^{mn}=\frac12
K^{m}{}_{AB}(y)K^{n}{}_{CD}(y)\left( u_{\underline{a}\underline{b}}{}^{AB}\,u_{\underline{a}\underline{b}}{}^{CD} 
+v^{\underline{a}\underline{b}\,AB}\,  v^{\underline{a}\underline{b}\,CD}\right)(x),
\end{eqnarray}
and
\begin{eqnarray}
A_{mnq}=\frac{\sqrt2}{4}~ \Delta~ g_{pq}(x,y)~ K_{mn}{}^{AB}(y)K^{q}{}_{CD}(y)\;\left(u^{\underline{a}\underline{b}}{}_{AB}
\,v^{\underline{a}\underline{b}\,CD}+ v_{\underline{a}\underline{b}\,AB}\,u_{\underline{a}\underline{b}}{}^{CD}\right)(x).
\end{eqnarray}
These expressions exactly agree with the ones provided in (\ref{MAmixframe}) and (\ref{FAmixframe}).
This is easily checked by using the Sp(56,$\mathbb R)$ condition of the SL(8,${\mathbb R})$ frame, $U_{SL_8}^T\Omega U_{SL_8}=\Omega$, implying
\begin{equation}
	\begin{array}{rcl}
		u_{ab}{}^{AB}v^{ab\,CD}-u_{ab}{}^{CD}v^{ab\,AB}&=&0\,,\\[3mm]
		u^{ab}{}_{AB}u_{ab}{}^{CD}-v_{ab\, AB}v^{ab\,CD}&=&\delta_{AB}^{CD}\,.
	\end{array}
\end{equation}
These relations, together with the orthogonality condition (\ref{ortho}), lead to the equivalent expressions
\begin{eqnarray}
\Delta^{-1}g^{mn}=\frac12
K^{m}{}_{AB}K^{n}{}_{CD}\left( u_{\underline{a}\underline{b}}{}^{AB}-iv^{\underline{a}\underline{b}\,AB}\right)\left(u_{\underline{a}\underline{b}}{}^{CD} 
+i v^{\underline{a}\underline{b}\,CD}\right)
\end{eqnarray}
and
\begin{eqnarray}
A_{mnq}=\frac{\sqrt2}{4}~ \Delta~ g_{pq}~ K_{mn}{}^{AB} K^{q}{}_{CD}\;\left( u^{\underline{a}\underline{b}}{}_{AB}+i\, v_{\underline{a}\underline{b}\,AB}\right)\left(u_{\underline{a}\underline{b}}{}^{CD} 
+i\, v_{\underline{a}\underline{b}\,AB}\right).
\end{eqnarray}
Finally, one can change the local indices $\underline{a}\underline{b}\rightarrow ij$ from $SL(8)$ to the $SU(8)$ basis to get the desired result.



\section{Outlook} 
\label{sec:outlook}

In this work we provided a new ansatz for the full uplift of the vacua of maximal gauged supergravity with non-compact gauge groups SO($p,q$) and tested it against the 11-dimensional equations of motion for all the known de Sitter vacua of these models.
This is still far from a complete non-linear ansatz for the metric and 3-form of 11-dimensional supergravity that includes the uplift of all the other fields, but it provides an interesting and successful intermediate step towards it.
Although one could be worried that the nature of the non-compact gauge group may create problems or pathologies in the uplifting procedure, we expect that the completion of our ansatz with the vector fields gauging the SO($p,q$) group will produce a consistent set of 11-dimensional tensors satisfying the 11-dimensional equations of motion.
It has already been checked in the past that there are efficient procedures to produce consistent truncations of higher-dimensional theories leading to lower-dimensional models with non-compact gauge groups, also in the case of hyperboloids as internal manifolds \cite{Cvetic:2004km}.
We expect that a generalization of such reduction procedures could be applied to our specific case.

An alternative way of deriving 4-dimensional gaugings of maximal supergravities uses a generalized form of the Scherk--Schwarz reduction procedure applied to Exceptional Generalized Geometry.
We showed that such a procedure gives the correct ansatz for the metric and 3-form field as follows from the construction of the generalized metric
\begin{eqnarray}
{\mathcal H}^{{\mathbb M}{\mathbb N}}(x,y)=E_{\mathbb A}{}^{\mathbb M}(y) {\mathcal M}^{{\mathbb A}{\mathbb B}}(x) 
E_{\mathbb B}{}^{\mathbb N}(x) .
\end{eqnarray}
While the construction of the generalized vielbein involves the use of Killing tensors for the space with $p,q$ signature, this procedure correctly reproduces euclidean geometries, because the scalar-dependent matrix ${\mathcal M}^{{\mathbb A}{\mathbb B}}(x)$ is positive definite.
Actually, the fact that the final metric depends on the contraction of the generalized vielbeins with a positive definite matrix ${\cal M}$, implies that at any vacuum the SO($p,q$) gauge group is broken to a subgroup of its maximal compact subgroup SO$(p)\times $ SO$(q)$. 

Generalized Scherk--Schwarz reductions have also been shown to be a powerful technique to describe lower dimensional field theories. 
In the context of Double Field Theory, they naturally lead to the electric sector of half maximal supergravity in 4 dimensions \cite{Aldazabal:2011nj} and these reductions also shed light on the higher dimensional interpretation of theories with non geometric fluxes (see $e.g.$ \cite{Dibitetto:2012rk} for an explicit construction of a solution with relaxed section condition in a particular reduction to 7 dimensions). 
A generalization of \cite{Aldazabal:2011nj} in the context of EFT with a 7 dimensional internal space is naturally connected to maximal supergravity in 4 dimensions, and our results show that not only the SO(8) but also the SO$(p,q)$ gauged supergravities can be interpreted in this frame. 
In this line of research it is interesting to explore if relaxing the section condition can lead to an uplift of the still misterious dyonic gauged supergravities of \cite{Dall'Agata:2012bb}.

\bigskip

\textbf{Note added:} While this work was under completion the paper \cite{Hohm:2014qga} appeared in the ArXiv, which discusses generalized Scherk--Schwarz reduction ans\"atze for Exceptional Field Theory on hyperboloids and therefore provides a different realization of the ideas in section~\ref{sec:hyperbloids_as_generalized_twisted_tori}.


\medskip 
\section*{Acknowledgments.}

\noindent We are glad to thank D.~Cassani, S.~Giusto, H.~Nicolai, E.~Plauschinn and A.~Rosabal for useful comments. 
This work was supported in part by the ERC Advanced Grants no.226455 (SUPERFIELDS) and by the MIUR grant RBFR10QS5J.

\end{document}